\documentclass[twocolumn,preprintnumbers,amsmath,amssymb,pra]{revtex4}

\pdfoutput=1
\usepackage{graphicx}
\usepackage{dcolumn}
\usepackage{bm}

\begin{document}


\title{Prospects for Doppler cooling of three-electronic-level molecules}

\author{J.H.V. Nguyen}
\author{B. Odom}
\affiliation{Department of Physics and Astronomy, Northwestern University, 2145 Sheridan Road,
Evanston, Illinois, USA 60208}

\date{\today}

\begin{abstract}
\noindent Analogous to the extension of laser cooling techniques from two-level to three-level atoms, Doppler cooling of molecules with an intermediate electronic state is considered.  In particular, we use a rate-equation approach to simulate cooling of $\mathrm{SiO}^{+}$, in which population buildup in the intermediate state is prevented by its short lifetime.  We determine that Doppler cooling of $\mathrm{SiO}^{+}$ can be accomplished without optically repumping from the intermediate state, at the cost of causing undesirable parity flips and rotational diffusion.  Since the necessary repumping would require a large number of continuous-wave lasers, optical pulse shaping of a femtosecond laser is proposed as an attractive alternative.  Other candidate three-electron-level molecules are also discussed.
\end{abstract}

\maketitle

\section{\label{sec:introduction}Introduction}

Cold molecule spectroscopy is a promising approach in searches for time variation of fundamental constants~\cite{PhysRevLett.99.150801, PhysRevLett.100.043201, PhysRevLett.100.043202, NewJPhys.11.055048}, parity violation in molecules~\cite{PhysRevLett.100.023003, 2010arXiv1007.1788I}, and time-reversal symmetry violation~\cite{JPhysB.43.074007}. Furthermore, cold polar molecules have been proposed for quantum computing~\cite{PhysRevLett.88.067901, 2009arXiv0903.3552S} and for long-range interaction studies of polar degenerate gases~\cite{Science.322.231}.  A review on recent developments in cold molecule technology  highlights these points, along with the difficulties associated with producing cold molecules~\cite{NewJPhys.11.055049}.

Translational cooling schemes have received significant attention~\cite{JChemPhys.104.9689, EurPhysJD.31.395, PhysRevLett.101.243002, PhysRevLett.103.223001, PhysRevA.77.023402, 2010arXiv1007.1788I}, and the DeMille group has recently demonstrated transverse Doppler cooling of $\mathrm{SrF}$~\cite{PhysRevLett.103.223001, Nature.467.820}.  To date, proposed schemes have been restricted to two-electronic-state molecules with highly diagonal Franck-Condon factors (FCFs), which are molecular equivalents of the simplest Doppler-cooling case: two-level atoms. (The FCF matrix describes the coupling of different vibrational levels in an electronic transition, with diagonal FCFs allowing many absorption-emission cycles without vibrational repumping.)  A number of two-electronic-state molecules have been proposed~\cite{EurPhysJD.31.395, PhysRevLett.101.243002, our.paper}; however, many potential candidates have an excited state whose lowest level lies \emph{above} the dissociation limit of the ground state~\cite{our.paper}.  Molecules having this characteristic suffer from predissociation~\cite{Herzberg}, in which coupling between the excited state and an unbound level of the ground state leads to dissociation on time scales which can be problematic for Doppler cooling~\cite{our.paper}.  

Interestingly, we have found a number of molecular ions with highly diagonal FCFs, which would not suffer from predissociation; however, all identified species have an off-diagonal intermediate electronic state.  Here, we build on existing work with two-electronic-state molecules and propose Doppler cooling of three-electronic-state molecules, potentially introducing a new class of coolable molecules.  This development somewhat parallels the situation for laser-coolable atoms, in which repumping out of a low-lying $D$ state considerably increased the list of coolable atomic candidates.

One possible solution in dealing with the intervening molecular electronic state is to increase the number of repump lasers, in order to return population to the cycling transition.  However, this approach would require a large number of repump wavelengths, because the molecule can decay into multiple intervening-state vibrational levels with significant probability (since the FCFs between the excited and intervening state are generally not diagonal).  Instead, we consider molecules which decay rapidly back into the ground state from the intermediate state.  If the decay from the intermediate state is sufficiently fast, then a repump is not necessarily required, particularly for the case of ion-trap experiments in which long confinement times allow one to wait for spontaneous decay from a dark state.  Unfortunately, spontaneous decay from the intervening state into the ground state results in diffusion of both parity and total angular momentum; however, the additional states can be repumped by using a pulse-shaped femtosecond laser, increasing the number of addressable transitions without a significant increase in the total number of lasers.

\section{\label{subsection:three-state-general}General considerations}

We require some number of repump lasers in order to scatter the $10^{4}$-$10^{6}$ photons required to reach the Doppler limit. The total number of lasers needed to pump the molecule back into the cycling transition depends on two main factors: the FCFs, which determine the magnitude of vibrational branching~\cite{EurPhysJD.31.395, PhysRevLett.103.223001, our.paper}, and the vibrational relaxation rate, which determines the magnitude of parity and total angular momentum diffusion~\cite{our.paper}.

\begin{figure}[t]
\begin{center}
\includegraphics[width=\columnwidth]{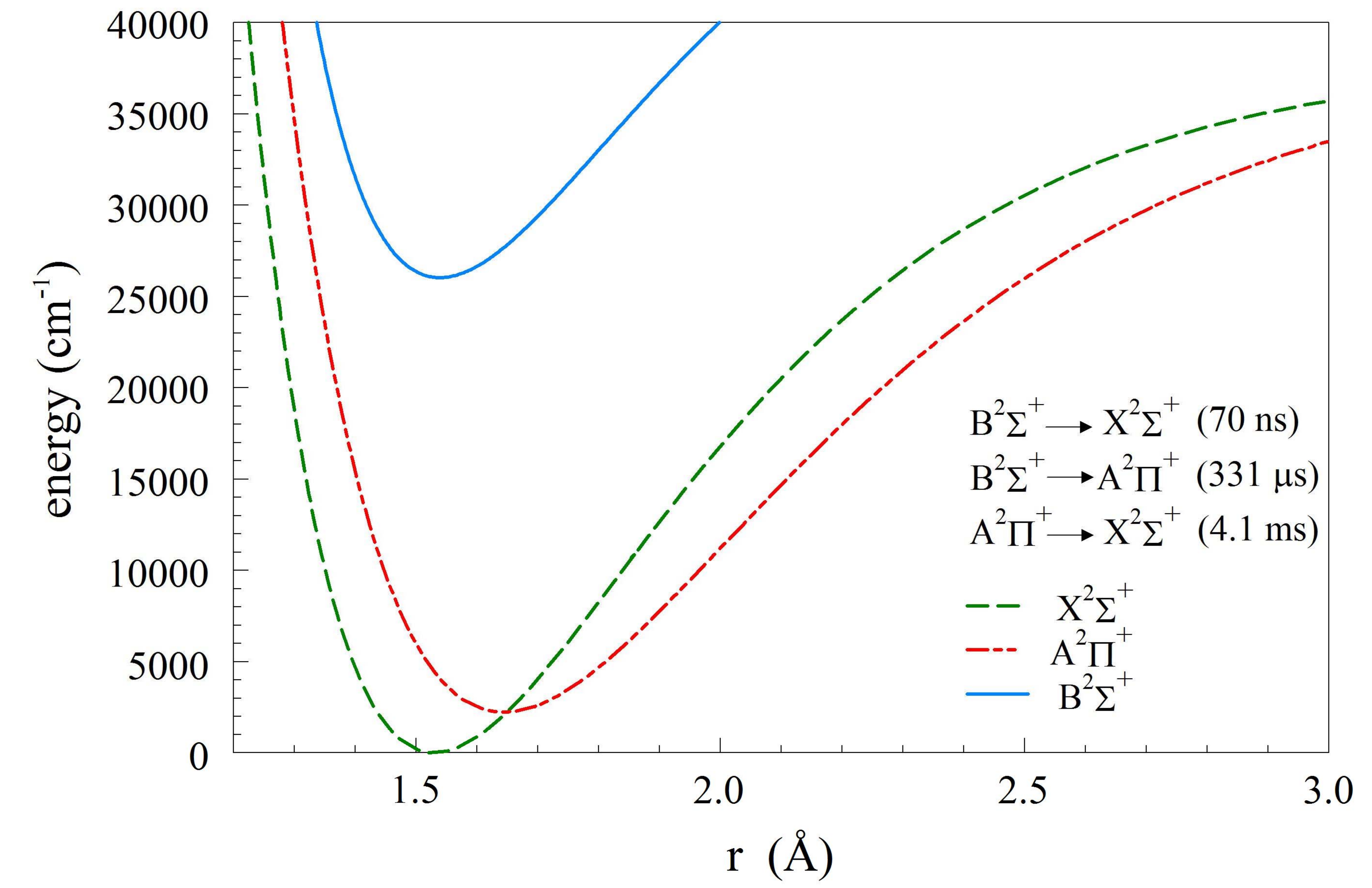}
\caption{\label{fig:SiO_PEC} (Color online) Potential energy curves generated from the molecular constants of Cai and Francois~\cite{JMolSpec.197.12}.  Lifetimes correspond to the net decay out of the $v=0$ level of the excited state.}
\end{center}
\end{figure}

\begin{figure}[b]
\begin{center}
\includegraphics[width=\columnwidth]{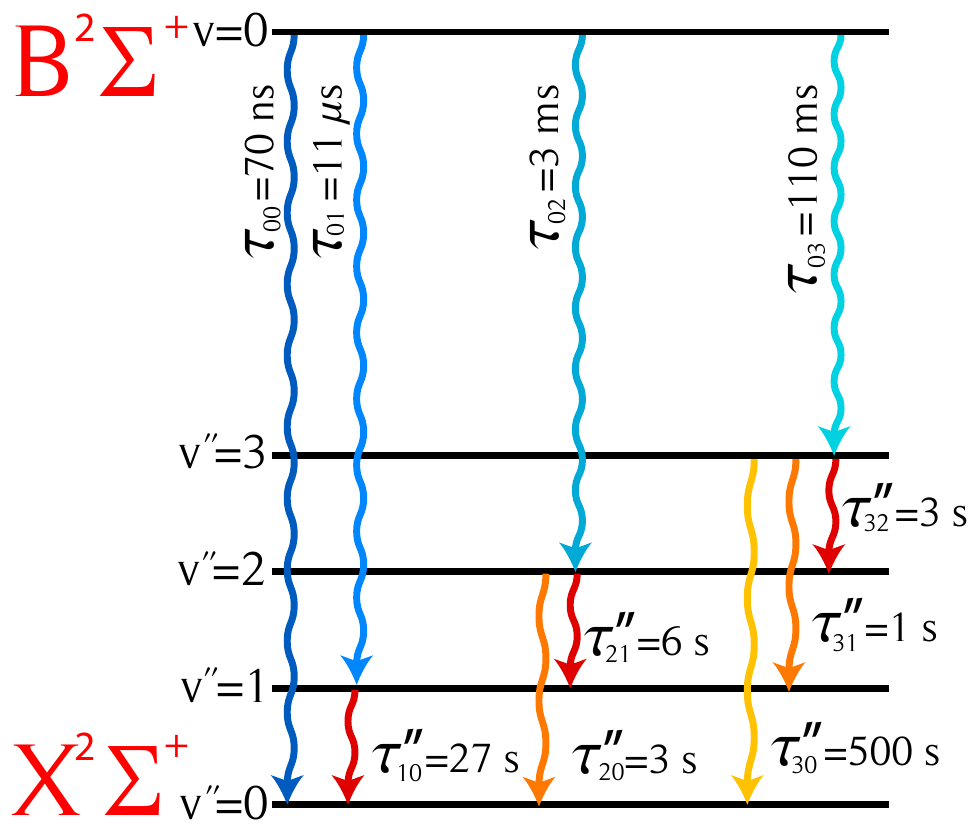}
\caption{\label{fig:simple_two_state} (Color online) Simplified energy level diagram for $\mathrm{SiO}^{+}$, omitting rotational substructure and the $\mathrm{A}^{2}\Pi$ state.  Electronic decay from the excited state is shown, as is vibrational decay within the ground electronic state (energy not to scale).}
\end{center}
\end{figure}

\begin{figure}[b]
\begin{center}
\includegraphics[width=\columnwidth]{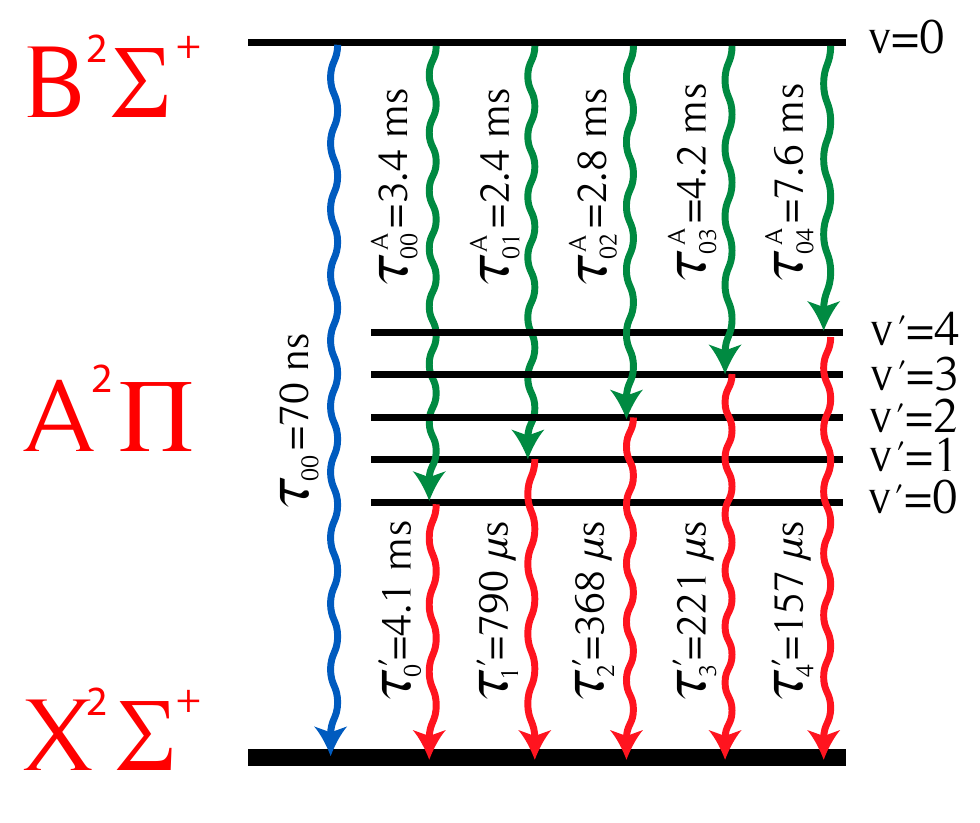}
\caption{\label{fig:simple_three_state} (Color online) Simplified energy-level diagram for $\mathrm{SiO}^{+}$, omitting the rotational substructure for all electronic states and  the vibrational substructure of the $X$ state.  Electronic decay into each vibrational level of the $A$ state is shown along with the net decay out of the level (energy not to scale).}
\end{center}
\end{figure}

As an example, we consider $\mathrm{SiO}^{+}$ (see Fig.~\ref{fig:SiO_PEC}), whose $\mathrm{X}{^{2}\Sigma^{+}}$ and $\mathrm{B}{^{2}\Sigma^{+}}$ states have highly diagonal FCFs (see Fig.~\ref{fig:simple_two_state}).  $\mathrm{SiO}^{+}$ also has an intervening $\mathrm{A}{^{2}\Pi}$ state, which lies $2000 \ \mathrm{cm}^{-1}$ ($60\ \mathrm{THz}$) above the ground state.  One would naively expect that depopulating the $A$ state would require additional repump lasers. However, $\mathrm{SiO}^{+}$ is representative of a class of molecules for which spontaneous decays from the intermediate state to the ground state are on the same time scale as the intermediate state is populated (see Fig.~\ref{fig:simple_three_state}).

Since the $A$ state in $\mathrm{SiO}^{+}$ lies only $2000\ \mathrm{cm}^{-1}$ ($60\ \mathrm{THz}$) above the $X$ state, one might expect that the $B \longrightarrow X$ and $B \longrightarrow A$ transitions would have similar spontaneous decay rates, since the transition rate is proportional to the cube of the transition frequency.  However, the transition moment for $B \longrightarrow A$ is surprisingly small~\cite{JMolSpec.197.12}, while the transition moment for $A \longrightarrow X$ transitions is relatively large~\cite{JMolSpec.197.12}.  The net result is that population buildup in the $A$ state is very slow. 

\begin{figure}[t]
\begin{center}
\includegraphics[width=\columnwidth]{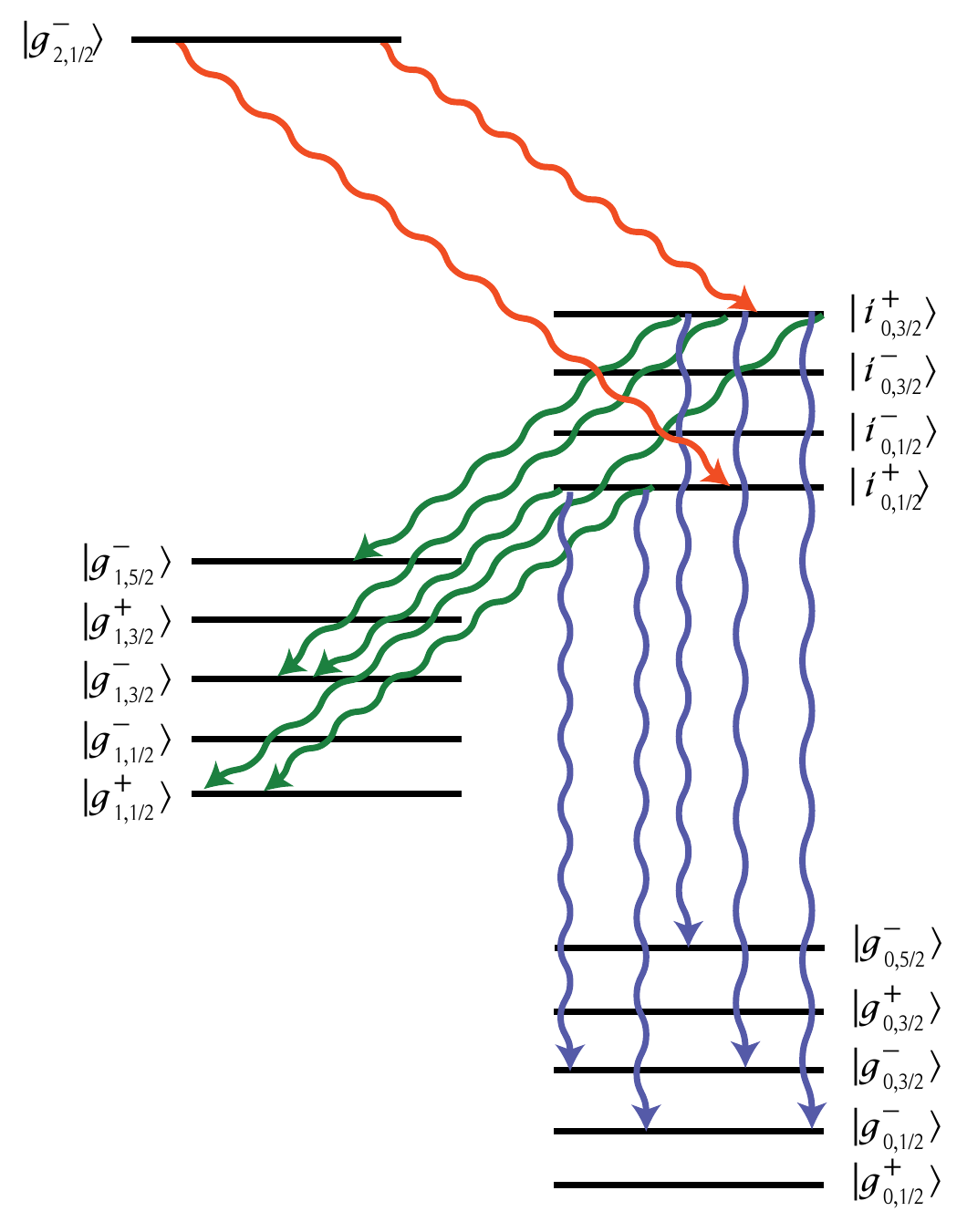}
\caption{\label{fig:SiO_intermediate_decay} (Color online) Simplified energy level diagram showing decay to the $|g^{P''}_{0,J''}\rangle$ states through the intermediate $|i^{P'}_{v',J'}\rangle$ states (energy not to scale).  Vibrational decay within the $X$ state and decay from the $|g^{P''}_{v>2,J''}\rangle$ states are not drawn.}
\end{center}
\end{figure}

\begin{figure}[t]
\begin{center}
\includegraphics[width=\columnwidth]{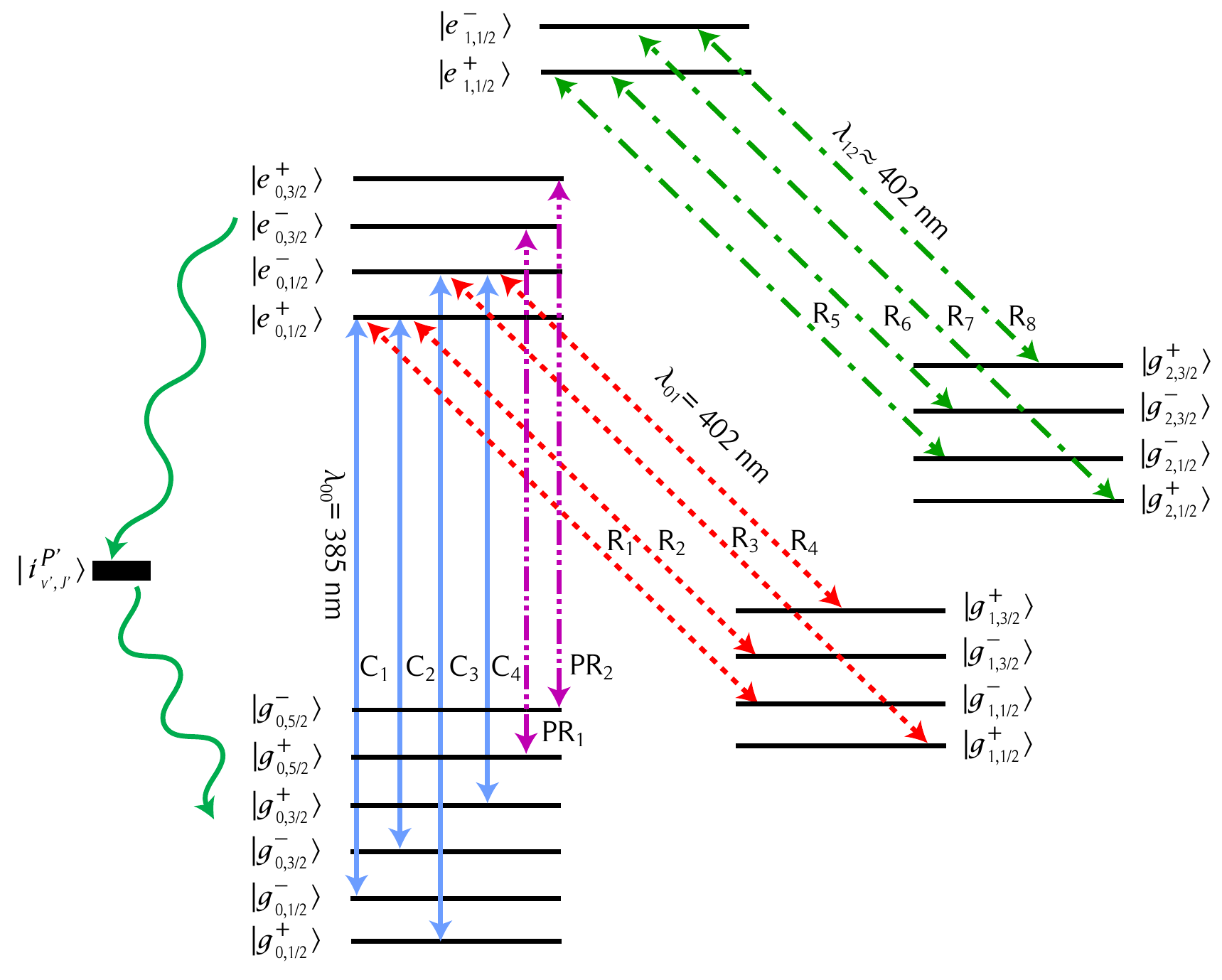}
\caption{\label{fig:lasers_three_state} (Color online) Simplified energy level diagram showing the cooling and repumping transitions for $\mathrm{SiO}^{+}$ with rotational substructure and both parities included (energy not to scale).  Transitions with different colors correspond to the addition of different lasers to the simulation, shown in Fig.~\ref{fig:three-state-solution}.}
\end{center}
\end{figure}

The low-lying $A$ state also provides a mechanism which increases the vibrational relaxation rate within the $X$ state.  This point becomes apparent from considering the $v'=0$ level of the $A$ state, which lies just below the $v''=2$ level of the $X$ state (see Fig.~\ref{fig:SiO_intermediate_decay}).  We denote states belonging to the $A$ state with $|i^{P'}_{v',J'}\rangle$.  The $|g^{P''}_{v''\geq 2,J''}\rangle \longleftrightarrow |i^{P'}_{v',J'}\rangle \longrightarrow |g^{P''}_{0,J''}\rangle$ decay channel provides a faster path to the $|g^{P''}_{0,J''}\rangle$ states than does direct vibrational decay within the $X$ manifold.  This rapid decay channel results in parity flips, requiring additional cycling and repumping lasers, as shown in Fig.~\ref{fig:lasers_three_state}.  Additionally, from the selection rule $\Delta J = 0, \pm 1$ it is clear that $B\longrightarrow A \longrightarrow X$ decay results in rotational diffusion, with population building up in the  $|g^{\pm}_{v'',J>3/2}\rangle$ states.

If the $A$ state were not present, $\mathrm{SiO}^{+}$ would behave similarly to $\mathrm{SrF}$ requiring repumping of a few vibrational levels of only one parity, and no rotational repumping.  However, the addition of the $A$ state effectively reduces the $X$ state vibrational lifetimes, resulting in three significant changes to the repumping and cycling requirements.  First, the even parity states $|g^{+}_{0,1/2}\rangle$ and $|g^{+}_{0,3/2}\rangle$ gain appreciable population and need to be repumped.   The $|g^{-}_{0,1/2}\rangle$ and $|g^{-}_{0,3/2}\rangle$ states can be driven by the same cw source, but the $|g^{+}_{0,1/2}\rangle$ and $|g^{+}_{0,3/2}\rangle$ are split by $129\ \mathrm{GHz}$, too large a shift for typical acousto-optic or electro-optic modulators.  Thus, the $|g^{+}_{0,1/2}\rangle \longrightarrow |e^{-}_{0,1/2}\rangle$ and $|g^{+}_{0,3/2}\rangle \longrightarrow |e^{-}_{0,1/2}\rangle$ must be driven by separate cw sources.  Second, each repumped vibrational manifold ($v''>0$) requires a total of four repumping wavelengths (see Fig.~\ref{fig:lasers_three_state}), of which only two can be derived from the same laser source.  Third, rotational repumping of the $|g^{\pm}_{v'',J''>3/2}\rangle$ states is required.  This rotational repumping can be accomplished by driving the $|g^{\pm}_{0,J''>3/2}\rangle \longleftrightarrow |e^{\mp}_{0,J''-1}\rangle$ transition ($P$-branch pumping) and $|g^{\pm}_{v''>0,J''>3/2}\rangle \longleftrightarrow |e^{\mp}_{v''-1,J''-1}\rangle$; however, each pumped rotational level requires a different cw source, since the rotational splitting is $0.7\ \mathrm{cm}^{-1}$ ($21\ \mathrm{GHz}$).

The actual number of driven transitions is determined in Sec.~\ref{sec:rate_equations}.  However, it is clear that the total number of required cw lasers quickly becomes unmanageable.  

An attractive alternative to using cw lasers for repumping is to use a broadband source, such as a femtosecond laser.  Femtosecond-pulsed vibrational repumping has been successfully demonstrated by Viteau \emph{et al.}, with a pulse-shaping resolution of $1\ \mathrm{cm}^{-1}$ ($30\ \mathrm{GHz}$)~\cite{Science.321.232}.  With the current resolution, this technique could be applied to rotational repumping of the two-electronic-level hydrides~\cite{our.paper}; applying this technique to $\mathrm{SiO}^{+}$ would require a resolution of better than $0.7 \ \mathrm{cm}^{-1}$ ($21\ \mathrm{GHz}$), which should be achievable.  A typical $\mathrm{SiO}^{+}$ repumped transition (e.g., $v''=1$) has a lifetime of $\tau_{01}=11\ \mathrm{\mu s}$, so the corresponding saturation intensity is $32\ \mathrm{\mu W / cm}^{2}$.  If the frequency-doubled output of a commercially available femtosecond laser ($P_{L} \approx 500\ \mathrm{mW})$, evenly distributed across its bandwidth ($\Delta_{L}\approx100\ \mathrm{cm}^{-1}$), is used for repumping, a $200\ \mathrm{\mu m}$ beam waist is sufficient to saturate the transition.  In fact, we can take advantage of the comb nature of the femtosecond laser and use a controlled sweep, allowing us to periodically saturate the repumping transition with a larger beam waist.  We note that for $\mathrm{SiO}^{+}$, the $P$ branch is well separated from the $R$ branch~\cite{AppSurfSci.197.202}, reducing the complexity of the required spectral filtering. One could imagine driving the cooling transitions with cw lasers and the repump transitions with one or two pulse-shaped lasers.

\section{\label{sec:rate_equations} Rate-equation simulation for $\mathrm{SiO}^{+}$}

The population dynamics of cooling are determined using a rate-equation approach, similar to that used to model population dynamics for internal cooling schemes~\cite{NatPhys.6.271, NatPhys.6.275}.  This approach allows us to demonstrate that population does not build in the intermediate state and also allows us to find the number of Doppler-cooling photons scattered by the cycling transition.  

Specifically, we solve

\begin{equation}
	\label{eq:rates_eq}
	\frac{d\bold{P}}{dt}=\bold{M}\bold{P},
\end{equation}

\noindent where $\bold{P}$ is a vector of $N$ energy-ordered elements, each corresponding to a different rovibrational state, and $\bold{M}$ is an $N\times N$ coupling matrix whose elements consist of Einstein $\mathcal{A}$ and $\mathcal{B}$ coefficients connecting the different rovibrational states.  Explicitly, the $i^{th}$ element is:

\begin{eqnarray}
	\label{eq:rates-ith-element}
	\frac{dP_{i}}{dt}&=& -\sum_{j=1}^{j=i-1}{\mathcal{A}_{ij}P_{i}}+\sum_{j=i+1}^{j=N}{\mathcal{A}_{ji}P_{j}} \\
	& & \nonumber -\sum_{j=1}^{j=i-1}{\mathcal{B}_{ij}E(\omega_{ij})P_{i}}-\sum_{j=i+1}^{j=N}{\mathcal{B}_{ij}E(\omega_{ij})P_{i}} \\
	& & \nonumber +\sum_{j=1}^{j=i-1}{\mathcal{B}_{ji}E(\omega_{ji})P_{j}}+\sum_{j=i+1}^{j=N}{\mathcal{B}_{ji}E(\omega_{ji})P_{j}} .
\end{eqnarray}

\noindent The $i^{th}$ and $j^{th}$ states are connected by Einstein coefficients denoted by $\mathcal{A}_{ij}$, $\mathcal{B}_{ij}$, and $\mathcal{B}_{ji}$ which correspond to spontaneous emission, stimulated emission, and absorption, respectively, and $E(\omega_{ij})$ is the spectral energy density at a given frequency $\omega_{ij}$.  The Einstein coefficients connecting different states are calculated using Level 8.0~\cite{level8.0}.  The energy separations were determined from Ref.~\cite{JMolSpec.169.364}, and potential energy curves and dipole transition moment curves were obtained from Refs.~\cite{JMolSpec.197.12, ChemPhys.73.169}.

In the three-electronic-level simulation, we include the $|g^{\pm}_{v''\leq 7, J''\leq13/2}\rangle$, the $|i^{\pm}_{v'\leq 7, J'\leq13/2}\rangle$, and the $|e^{\pm}_{v\leq 2, J\leq11/2}\rangle$ states for a total of $329$ states.  Transitions between electronic states are included, whereas rovibrational transitions within an electronic state are neglected, since these events are rare on cooling time scales.  However, the effective vibrational-relaxation rate for the $X$ state is reduced since the intervening $A$ state provides a fast-decay mechanism through the $X \longrightarrow A \longleftrightarrow X$ decay channel, which is included in the simulation.

\begin{figure}[tbf]
\begin{center}
\includegraphics[width=\columnwidth]{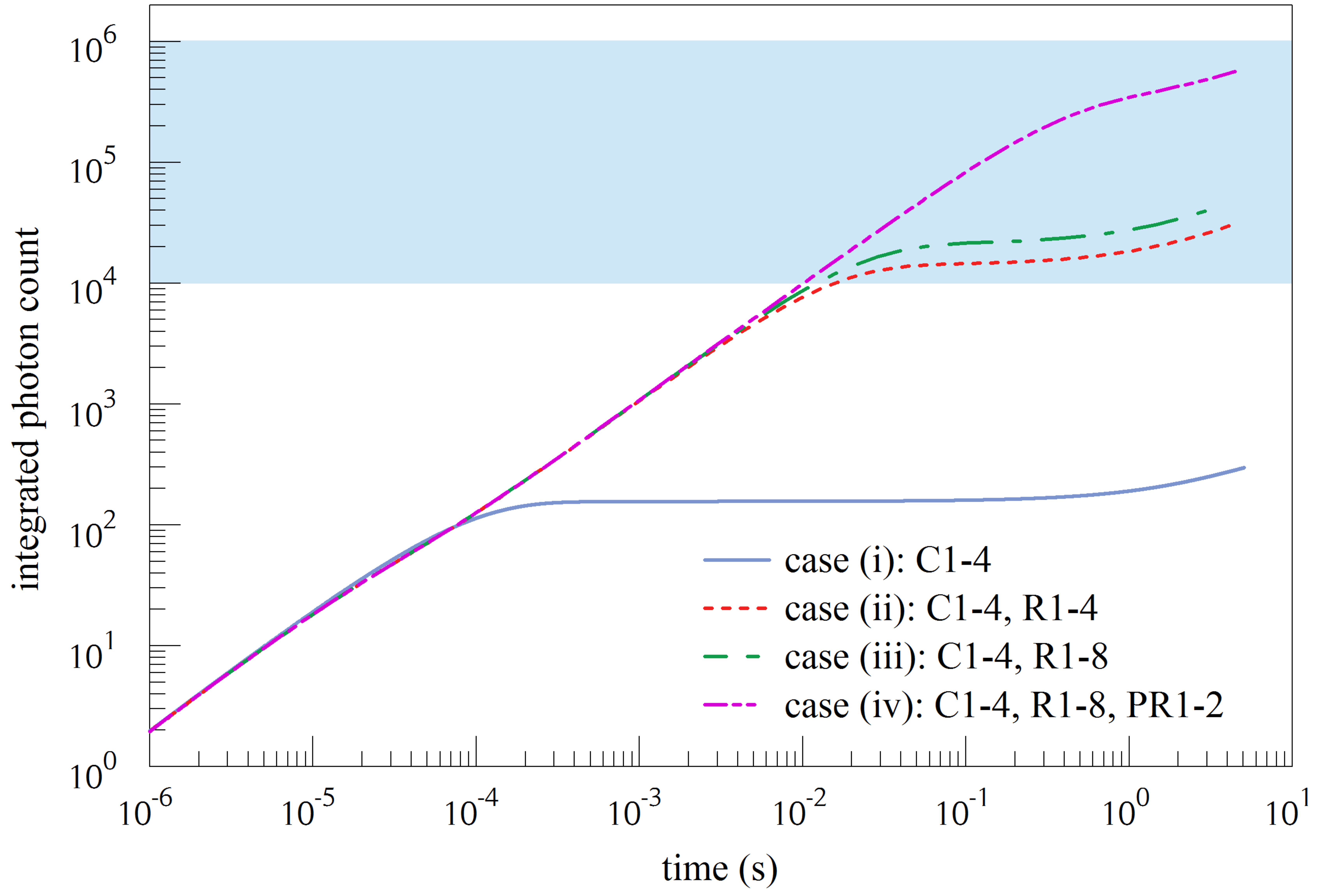}
\caption{\label{fig:three-state-solution} (Color online) Total number of photons scattered for different laser configurations.  The shaded band corresponds to cooling to the Doppler limit from room temperature.  The upper limit corresponds a laser detuning fixed at half the natural line width, and the lower limit corresponds to a variable detuning that follows the temperature-dependent Doppler width.  Cooling (C$x$) and repumping (R$x$) wavelengths are shown in Fig.~\ref{fig:lasers_three_state}. From the plot, we see that we require a vibrational repump for the $v'' \leq 2$ levels as well as a rotational repump for $J''=\pm 5/2$. }
\end{center}
\end{figure}

Population is initially set to the $|g^{-}_{0,1/2}\rangle$ state, and the simulation is repeated for different laser configurations. In Fig.~\ref{fig:three-state-solution}, we show the number of scattered photons from the cycling transition.   In Case (i), we apply four wavelengths to drive the $|g^{\pm}_{0,1/2}\rangle\longleftrightarrow |e^{\mp}_{0,1/2}\rangle$ and $|g^{\pm}_{0,3/2}\rangle\longleftrightarrow |e^{\mp}_{0,1/2}\rangle$ cycling transitions (lines C$_{1}$-C$_{4}$ in Fig.~\ref{fig:lasers_three_state}).  Parity flips resulting from the intervening $A$ state require cycling both parities. Population builds up in the $|g^{\pm}_{1,1/2}\rangle$ and $|g^{\pm}_{1,1/2}\rangle$ states.

In Case (ii), we apply an additional four wavelengths to drive the $|g^{\pm}_{1,1/2}\rangle\longleftrightarrow |e^{\mp}_{0,1/2}\rangle$ and $|g^{\pm}_{1,3/2}\rangle\longleftrightarrow |e^{\mp}_{0,1/2}\rangle$ repumping transitions (lines R$_{1}$-R$_{4}$ in Fig.~\ref{fig:lasers_three_state}).   If the $A$ state were not present, population would primarily go dark in the $v''=2$ level since the $v''=1$ level is repumped, and because decay into the $v''=2$ level occurs at a faster rate than does decay into the $v''=3$ level (see Fig.~\ref{fig:simple_two_state}).   However, decay into multiple vibrational levels of the $A$ state occurs on the same timescale as does decay into the $v''=2$ level of the $X$ state (see Figs.~\ref{fig:simple_two_state} and \ref{fig:simple_three_state}).  Subsequent decay from the $A$ to $X$ state is fast, and the FCFs are such that population primarily returns to the $v''=0$ level, resulting in diffusion of parity and total angular momentum.  The resultant total photon count of $N\approx 3 \times 10^{4}$ photons is barely above the minimum required number of scattered photons.  

In Case (iii), we apply an additional four wavelengths to drive the $|g^{\pm}_{2,1/2}\rangle\longleftrightarrow |e^{\mp}_{1,1/2}\rangle$ and $|g^{\pm}_{2,3/2}\rangle\longleftrightarrow |e^{\mp}_{1,1/2}\rangle$ repumping transitions (lines R$_{5}$-R$_{8}$ in Fig.~\ref{fig:lasers_three_state}), resulting in a slight increase in the photon count ($N\approx7 \times 10^{4}$). 

In Case (iv), we apply an additional two wavelengths to drive the $|g^{\pm}_{0,5/2}\rangle\longleftrightarrow |e^{\mp}_{0,3/2}\rangle$ ($P$ branch) transitions (lines PR$_{1}$ and PR$_{2}$ in Fig.~\ref{fig:lasers_three_state}).  Now, population builds up in the $|g^{\pm}_{1,5/2}\rangle$ states, and the total number of scattering events is well above the minimum required for cooling to the Doppler limit ($N\approx 6\times 10^{5}$).

\section{\label{solution_discussion}Discussion}

\subsection{\label{fs-assisted-candidates}Femtosecond-assisted Doppler cooling candidates}

The approach of repumping by femtosecond laser pulse shaping increases the list of coolable molecules to include two-electronic-state hydrides~\cite{our.paper}, and certain molecules with a third intermediate state.  For example, a low-lying $A$ state also occurs in molecular ions: $\mathrm{CO}^{+}$~\cite{ChemPhys.130.361}, $\mathrm{GeO}^{+}$~\cite{ChemPhys.315.35}, $\mathrm{SnO}^{+}$~\cite{ChemPhysLett.368.465, JPhysChem.88.5759}, and $\mathrm{PbO}^{+}$~\cite{JPhysChem.88.5759}; and for neutral molecules: $\mathrm{BO}$~\cite{ChinOptLett.8.663}, $\mathrm{AlO}$~\cite{JMolStruct.458.61}, $\mathrm{GaO}$~\cite{JMolStruct.672.105}, $\mathrm{InO}$~\cite{MolPhys.89.13}, $\mathrm{TlO}$, and $\mathrm{N}_{2}^{+}$~\cite{JChemPhys.87.4716, JChemPhys.88.329}.  An intermediate state that has a different spin multiplicity (an $a$ state) that lies approximately equidistant to the excited and ground states also occurs in molecular ions: $\mathrm{CH}^{+}$~\cite{MolPhys.102.23}, $\mathrm{SiH}^{+}$~\cite{ChemPhsLett.237.204}, and $\mathrm{CCl}^{+}$~\cite{JChemPhys.78.7260}.  Two cases discussed further below are $\mathrm{CO}^{+}$, whose intermediate state lies between the $X$ and $B$ states, and $\mathrm{N}_{2}^{+}$, whose ungerade intermediate state is not dipole connected to the ungerade excited state.

In $\mathrm{CO}^{+}$,  the B$\longrightarrow$X transition has a rotationless radiative lifetime of $57.1 \ \mathrm{ns}$ (for $v=0$) with a transition frequency of $46,520\ \mathrm{cm}^{-1}$ ($215\ \mathrm{nm}$), and the A$\longrightarrow$X transition has a rotationless radiative lifetime of $4.03\ \mathrm{\mu s}$ (for $v=0$) with a transition frequency of $20,986\ \mathrm{cm}^{-1}\ (477\ \mathrm{nm})$~\cite{ChemPhys.130.361}.   The radiative lifetime for the B$\longrightarrow$A transition is $3.07\ \mathrm{\mu s}$, which we estimate by scaling the A$\longrightarrow$X lifetime by the transition frequency ($25,534\ \mathrm{cm}^{-1}\ (392\ \mathrm{nm})$), and by the transition moment which is $65 \%$ smaller in the B$\longrightarrow$A transition than the A$\longrightarrow$X, at the equilibrium bond length.  The short lifetime of the B$\longrightarrow$X transition, and the comparably long lifetimes of the B$\longrightarrow$A transition and the A$\longrightarrow$X transition make $\mathrm{CO}^{+}$ a potential cooling candidate, although a more careful analysis of transition rates would be required to determine the necessary repump transitions.

Another interesting potential candidate is $\mathrm{N}_{2}^{+}$, which has been loaded into an ion trap via state-selective photoionization~\cite{PhysRevLett.105.143001}.    Similar to $\mathrm{SiO}^{+}$ and $\mathrm{CO}^{+}$, the B$\longrightarrow$X transition of $\mathrm{N}_{2}^{+}$ has a short radiative lifetime ($59.1 \ \mathrm{ns}$ for $v=0$) with a transition frequency of $25,823 \ \mathrm{cm}^{-1}\ (387\ \mathrm{nm})$~\cite{JChemPhys.88.329}, and the $A \longrightarrow X$ transition has a radiative lifetime of $14\ \mathrm{ms}$ with a transition frequency of $9150\ \mathrm{cm}^{-1}$ ($1.1\ \mathrm{\mu m}$)~\cite{JChemPhys.87.4716}.    However, inversion symmetry prevents decay from the $B$ state ($^{2}\Sigma_{u}^{+}$) to the $A$ state ($^{2}\Pi_{u}$), so population does not appreciably build up in the $A$ state.  Also, since the molecule is homonuclear, there is no permanent dipole moment, so rovibrational transitions are dipole-forbidden.  Unfortunately, the FCFs of $\mathrm{N}_{2}^{+}$ are not as diagonal as they are for $\mathrm{SiO}^{+}$, so more vibrational repumping lasers would be required.

\subsection{\label{solution_doppler} Doppler cooling of SiO$^{+}$}

\begin{figure}[tbf]
\begin{center}
\includegraphics[width=\columnwidth]{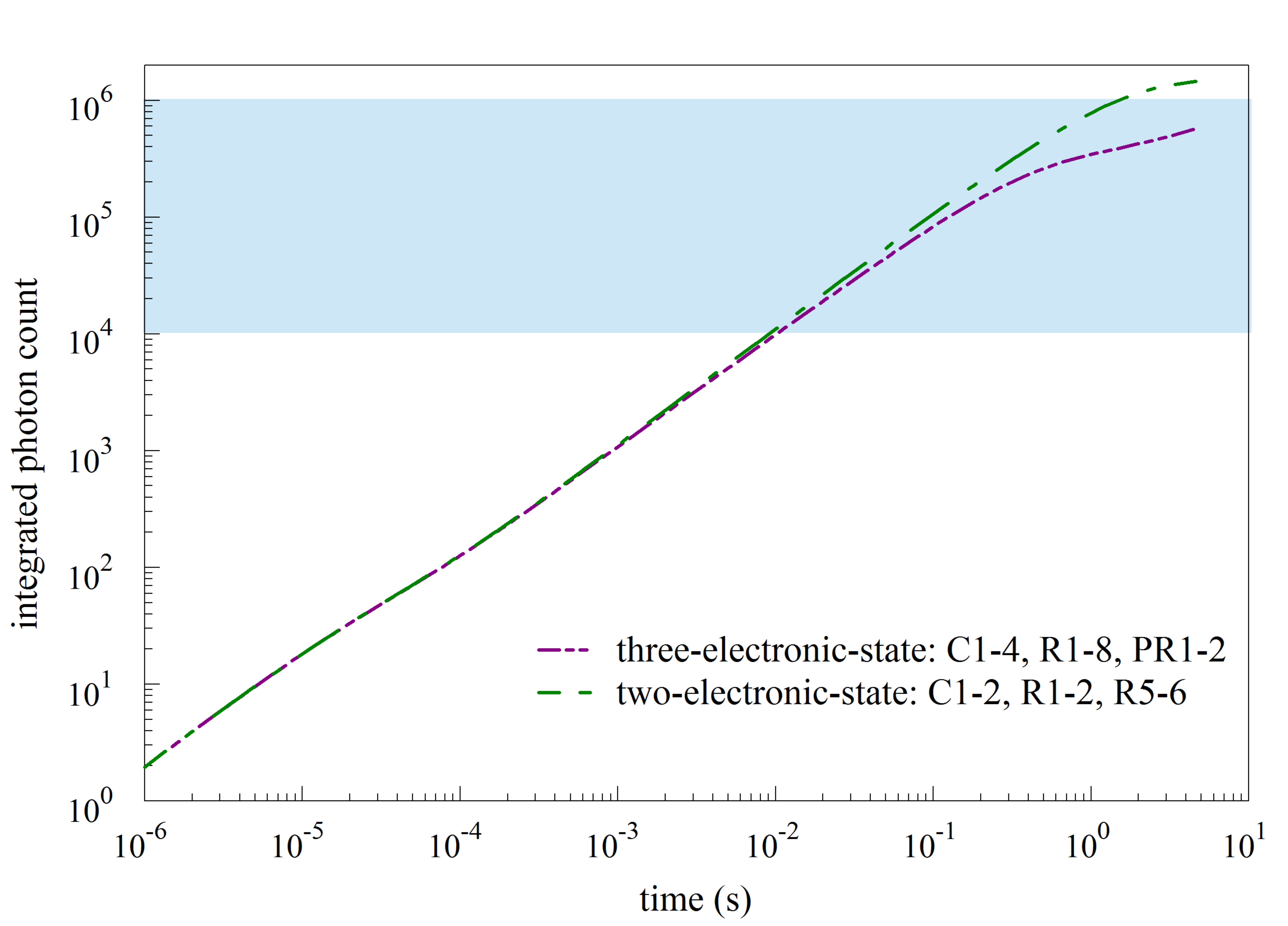}
\caption{\label{fig:photon_count_comp} (Color online) Total number of photons scattered, treating $\mathrm{SiO}^{+}$ alternatively as a two-electronic-state molecule (ignoring the $A$ state) and as a three-electronic-state molecule (full treatment).  The shaded band corresponds to cooling to the Doppler limit from room temperature for different laser detunings (similar to Fig.~\ref{fig:three-state-solution}).  Cooling (C$x$) and repumping (R$x$) wavelengths are shown in Fig.~\ref{fig:lasers_three_state}.}
\end{center}
\end{figure}

By modeling the population dynamics using a rate-equation approach, we determine that $\mathrm{SiO}^{+}$ is a promising candidate for Doppler cooling.  The FCFs between the ground and excited states are highly diagonal, with a $B$ state lifetime of $70 \ \mathrm{ns}$.  Decay from the intermediate $A$ state occurs on a time scale faster than decay into it, so the effect of the intervening state is negligible, except that it leads to parity and rotational diffusion.  Furthermore, the $B$ state minimum lies below the dissociation limit of the $X$ and $A$ states, so predissociation into the $X$ or $A$ continua~\cite{our.paper} does not occur.

We must also examine the higher lying electronic states of $\mathrm{SiO}^{+}$.  In our model, we assumed a three-electronic-state system.  In fact, there are higher lying electronic states which we may accidentally excite from the $B$ state.  Coupling to the higher states would be of concern if they are unbound or if they predissociate~\cite{our.paper}.  We are currently in the process of performing spectroscopy with $\mathrm{SiO}^{+}$ to determine if these electronic states are connected via two-photon transitions from $X \rightarrow B$ and subsequently from the $B$ state to a higher-lying state.

From the simulation for the three-electronic-state system, we find a total of four cycling transitions and ten repumping transitions are required.  The four cycling transitions require three separate lasers since the $|g^{-}_{0,1/2}\rangle$ and $|g^{-}_{0,3/2}\rangle$ states belong to the same spin-rotation-split pair ($\Delta \omega_{sr} \approx 2\pi \times 625\ \mathrm{MHz}$).  Of the ten repumping transitions, eight can be driven by the same femtosecond laser source: the $|g^{\pm}_{1,J''=1/2, 3/2}\rangle \longrightarrow |e^{\mp}_{0,1/2}\rangle$, and the $|g^{\pm}_{2,J''=1/2,3/2}\rangle \longrightarrow |e^{\mp}_{1,J'}\rangle$ transitions, which span a region of $41\ \mathrm{cm}^{-1}$ ($1.2\ \mathrm{THz}$).  The remaining two $P$-branch pumping transitions ($|g^{\pm}_{0,5/2}\rangle \longrightarrow |e^{\mp}_{0,5/2}\rangle$) need to be driven with a separate cw laser for each, or with another femtosecond laser. 

We also include a comparison of the number of scattered photons for $\mathrm{SiO}^{+}$ for Case (iv) of the three-electronic-state system, and a simulation where the intervening state was neglected (see Fig.~\ref{fig:photon_count_comp}).  Here, we see that the intervening $A$ state changes the nature of the repumping scheme; however, the $A$ state itself need not be repumped.  At $t\approx10^{-2}\ \mathrm{s}$ we notice a reduction in counts for the three-electronic-state system, which is due to population building up in the $|g^{\pm}_{1,5/2}\rangle$ states.  These states can be repumped using the same femtosecond laser present in this scheme, since the corresponding repump transitions fall within the laser bandwidth.

Using a femtosecond laser does require additional consideration.  Although the overall envelope of the laser is broad, the individual frequency components are comblike.  For a typical femtosecond laser, the frequency components are separated by $80\ \mathrm{MHz}$, so it is likely that a repumping transition will lie within the space between two frequency-comb lines.  However, various modulation techniques, either internal or external to the laser, can be used to span the appropriate repumping transitions.

\section{\label{sec:conclusion} Conclusion}

Similarly to how a low-lying $D$ state does not preclude Doppler cooling of three-level atoms, an intermediate electronic state does not preclude Doppler cooling of three-electronic-level molecules.  Using a rate-equation approach to model the population dynamics of $\mathrm{SiO}^{+}$ we determine that this species is a candidate for Doppler cooling of three-electronic-level molecules.  The $X$ and $B$ states have highly diagonal FCFs, and population does not remain trapped within the low-lying $A$ state due to $A \longrightarrow X$ decay.  As compared with a polar two-electronic-level molecule of similar reduced mass, the total number of repumped transitions is higher as a result of diffusion into opposite parity and higher-lying rotational states; however, this complication does not result in a significant increase in the total number of lasers when a pulse-shaped femtosecond laser is used for repumping.  Alternatively, if cryogenic precooling is used for trapped ions~\cite{2009arXiv0903.3552S} or for neutral beams~\cite{JPhysB.43.074007, PhysRevLett.103.223001, Nature.467.820}, the number of photons scattered to reach the Doppler limit is reduced by up to nearly two orders of magnitude. In this case, Doppler cooling of two-electronic-level hydrides~\cite{our.paper} or three-electronic-level oxides (see Fig.~\ref{fig:three-state-solution}) requires fewer repumping lasers. 

For the case of molecular ions, sympathetic cooling by a cotrapped atomic species provides an alternate and quite general approach~\cite{NatPhys.6.271,NatPhys.6.275, PhysRevLett.105.143001} for achieving milliKelvin molecular ion temperatures.  However, direct
Doppler cooling of the molecular ion species removes the need for a sympathetic coolant ion, which would be desirable in cases in which the light driving the atomic-cycling transition changes the molecular internal state or results in photodissociation. Elimination of the sympathetic coolant ions might also be desirable for cases in which the presence of a second species could create unwanted complications (e.g., quantum information processing using trapped polar molecular ions coupled to external circuits~\cite{2009arXiv0903.3552S}).  Additionally, achieving the degree of photon cycling necessary for Doppler cooling would also allow single-molecule fluorescence imaging and nondestructive direct detection of trapped molecular species.

In conclusion, an intermediate electronic state is not prohibitive to Doppler cooling and does not require optical pumping when population rapidly decays out of that state.  We note that the requirements for repumping rates and intermediate-state relaxation rates are considerably relaxed in ion traps, since very long state-independent trapping times are routinely achieved. We have also suggested a number of other possible candidates which are similar to $\mathrm{SiO}^{+}$, as well as molecules in which the intermediate state is not dipole-connected because of spin or gerade/ungerade symmetry.

Currently, we are in the process of performing spectroscopy of $\mathrm{SiO}^{+}$ in order to make a first observation of electronic states above the $B$ state.  Once the energy for these transitions is confirmed, we plan on proceeding with an attempt to Doppler cool $\mathrm{SiO}^{+}$.

\begin{acknowledgments}
During preparation of this manuscript, we discovered that Wes Campbell was also thinking about Doppler cooling of $\mathrm{SiO}^{+}$.  We would like to thank Ken Brown, Eric Hudson, and Marcin Zi\'{o}\l kowski for helpful conversations.  This work was supported by the David and Lucile Packard Foundation (Grant No. 2009-34713) and the AFOSR (Grant No. FA9550-10-1-0221).
\end{acknowledgments}

\bibliography{SiO_plus_paper}

\begin{thebibliography}{41}
\expandafter\ifx\csname natexlab\endcsname\relax\def\natexlab#1{#1}\fi
\expandafter\ifx\csname bibnamefont\endcsname\relax
  \def\bibnamefont#1{#1}\fi
\expandafter\ifx\csname bibfnamefont\endcsname\relax
  \def\bibfnamefont#1{#1}\fi
\expandafter\ifx\csname citenamefont\endcsname\relax
  \def\citenamefont#1{#1}\fi
\expandafter\ifx\csname url\endcsname\relax
  \def\url#1{\texttt{#1}}\fi
\expandafter\ifx\csname urlprefix\endcsname\relax\def\urlprefix{URL }\fi
\providecommand{\bibinfo}[2]{#2}
\providecommand{\eprint}[2][]{\url{#2}}

\bibitem[{\citenamefont{Flambaum and Kozlov}(2007)}]{PhysRevLett.99.150801}
\bibinfo{author}{\bibfnamefont{V.~V.} \bibnamefont{Flambaum}} \bibnamefont{and}
  \bibinfo{author}{\bibfnamefont{M.~G.} \bibnamefont{Kozlov}},
  \bibinfo{journal}{Phys. Rev. Lett.} \textbf{\bibinfo{volume}{99}},
  \bibinfo{pages}{150801} (\bibinfo{year}{2007}).

\bibitem[{\citenamefont{Zelevinsky et~al.}(2008)\citenamefont{Zelevinsky,
  Kotochigova, and Ye}}]{PhysRevLett.100.043201}
\bibinfo{author}{\bibfnamefont{T.}~\bibnamefont{Zelevinsky}},
  \bibinfo{author}{\bibfnamefont{S.}~\bibnamefont{Kotochigova}},
  \bibnamefont{and} \bibinfo{author}{\bibfnamefont{J.}~\bibnamefont{Ye}},
  \bibinfo{journal}{Phys. Rev. Lett.} \textbf{\bibinfo{volume}{100}},
  \bibinfo{pages}{043201} (\bibinfo{year}{2008}).

\bibitem[{\citenamefont{DeMille
  et~al.}(2008{\natexlab{a}})\citenamefont{DeMille, Sainis, Sage, Bergeman,
  Kotochigova, and Tiesinga}}]{PhysRevLett.100.043202}
\bibinfo{author}{\bibfnamefont{D.}~\bibnamefont{DeMille}},
  \bibinfo{author}{\bibfnamefont{S.}~\bibnamefont{Sainis}},
  \bibinfo{author}{\bibfnamefont{J.}~\bibnamefont{Sage}},
  \bibinfo{author}{\bibfnamefont{T.}~\bibnamefont{Bergeman}},
  \bibinfo{author}{\bibfnamefont{S.}~\bibnamefont{Kotochigova}},
  \bibnamefont{and} \bibinfo{author}{\bibfnamefont{E.}~\bibnamefont{Tiesinga}},
  \bibinfo{journal}{Phys. Rev. Lett.} \textbf{\bibinfo{volume}{100}},
  \bibinfo{pages}{043202} (\bibinfo{year}{2008}{\natexlab{a}}).

\bibitem[{\citenamefont{Chin et~al.}(2009)\citenamefont{Chin, Flambaum, and
  Kozlov}}]{NewJPhys.11.055048}
\bibinfo{author}{\bibfnamefont{C.}~\bibnamefont{Chin}},
  \bibinfo{author}{\bibfnamefont{V.~V.} \bibnamefont{Flambaum}},
  \bibnamefont{and} \bibinfo{author}{\bibfnamefont{M.~G.}
  \bibnamefont{Kozlov}}, \bibinfo{journal}{New J. Phys.}
  \textbf{\bibinfo{volume}{11}}, \bibinfo{pages}{055048}
  (\bibinfo{year}{2009}).

\bibitem[{\citenamefont{DeMille
  et~al.}(2008{\natexlab{b}})\citenamefont{DeMille, Cahn, Murphree, Rahmlow,
  and Kozlov}}]{PhysRevLett.100.023003}
\bibinfo{author}{\bibfnamefont{D.}~\bibnamefont{DeMille}},
  \bibinfo{author}{\bibfnamefont{S.~B.} \bibnamefont{Cahn}},
  \bibinfo{author}{\bibfnamefont{D.}~\bibnamefont{Murphree}},
  \bibinfo{author}{\bibfnamefont{D.~A.} \bibnamefont{Rahmlow}},
  \bibnamefont{and} \bibinfo{author}{\bibfnamefont{M.~G.}
  \bibnamefont{Kozlov}}, \bibinfo{journal}{Phys. Rev. Lett.}
  \textbf{\bibinfo{volume}{100}}, \bibinfo{pages}{023003}
  (\bibinfo{year}{2008}{\natexlab{b}}).

\bibitem[{\citenamefont{{Isaev} et~al.}(2010)\citenamefont{{Isaev}, {Hoekstra},
  and {Berger}}}]{2010arXiv1007.1788I}
\bibinfo{author}{\bibfnamefont{T.~A.} \bibnamefont{{Isaev}}},
  \bibinfo{author}{\bibfnamefont{S.}~\bibnamefont{{Hoekstra}}},
  \bibnamefont{and} \bibinfo{author}{\bibfnamefont{R.}~\bibnamefont{{Berger}}},
  \bibinfo{journal}{Phys. Rev. A} \textbf{\bibinfo{volume}{82}},
  \bibinfo{pages}{052521} (\bibinfo{year}{2010}).

\bibitem[{\citenamefont{Vutha et~al.}(2010)\citenamefont{Vutha, Campbell,
  Gurevich, Hutzler, Parsons, Patterson, Petrik, Spaun, Doyle, Gabrielse
  et~al.}}]{JPhysB.43.074007}
\bibinfo{author}{\bibfnamefont{A.~C.} \bibnamefont{Vutha}},
  \bibinfo{author}{\bibfnamefont{W.~C.} \bibnamefont{Campbell}},
  \bibinfo{author}{\bibfnamefont{Y.~V.} \bibnamefont{Gurevich}},
  \bibinfo{author}{\bibfnamefont{N.~R.} \bibnamefont{Hutzler}},
  \bibinfo{author}{\bibfnamefont{M.}~\bibnamefont{Parsons}},
  \bibinfo{author}{\bibfnamefont{D.}~\bibnamefont{Patterson}},
  \bibinfo{author}{\bibfnamefont{E.}~\bibnamefont{Petrik}},
  \bibinfo{author}{\bibfnamefont{B.}~\bibnamefont{Spaun}},
  \bibinfo{author}{\bibfnamefont{J.~M.} \bibnamefont{Doyle}},
  \bibinfo{author}{\bibfnamefont{G.}~\bibnamefont{Gabrielse}},
  \bibnamefont{et~al.}, \bibinfo{journal}{J. Phys. B}
  \textbf{\bibinfo{volume}{43}}, \bibinfo{pages}{074007}
  (\bibinfo{year}{2010}).

\bibitem[{\citenamefont{DeMille}(2002)}]{PhysRevLett.88.067901}
\bibinfo{author}{\bibfnamefont{D.}~\bibnamefont{DeMille}},
  \bibinfo{journal}{Phys. Rev. Lett.} \textbf{\bibinfo{volume}{88}},
  \bibinfo{pages}{067901} (\bibinfo{year}{2002}).

\bibitem[{\citenamefont{Schuster et~al.}(2011)\citenamefont{Schuster, Bishop,
  Chuang, DeMille, and Schoelkopf}}]{2009arXiv0903.3552S}
\bibinfo{author}{\bibfnamefont{D.~I.} \bibnamefont{Schuster}},
  \bibinfo{author}{\bibfnamefont{L.~S.} \bibnamefont{Bishop}},
  \bibinfo{author}{\bibfnamefont{I.~L.} \bibnamefont{Chuang}},
  \bibinfo{author}{\bibfnamefont{D.}~\bibnamefont{DeMille}}, \bibnamefont{and}
  \bibinfo{author}{\bibfnamefont{R.~J.} \bibnamefont{Schoelkopf}},
  \bibinfo{journal}{Phys. Rev. A} \textbf{\bibinfo{volume}{83}},
  \bibinfo{pages}{012311} (\bibinfo{year}{2011}).

\bibitem[{\citenamefont{Ni et~al.}(2008)\citenamefont{Ni, Ospelkaus,
  de~Miranda, Pe'er, Neyenhuis, Zirbel, Kotochigova, Julienne, Jin, and
  Ye}}]{Science.322.231}
\bibinfo{author}{\bibfnamefont{K.-K.} \bibnamefont{Ni}},
  \bibinfo{author}{\bibfnamefont{S.}~\bibnamefont{Ospelkaus}},
  \bibinfo{author}{\bibfnamefont{M.~H.~G.} \bibnamefont{de~Miranda}},
  \bibinfo{author}{\bibfnamefont{A.}~\bibnamefont{Pe'er}},
  \bibinfo{author}{\bibfnamefont{B.}~\bibnamefont{Neyenhuis}},
  \bibinfo{author}{\bibfnamefont{J.~J.} \bibnamefont{Zirbel}},
  \bibinfo{author}{\bibfnamefont{S.}~\bibnamefont{Kotochigova}},
  \bibinfo{author}{\bibfnamefont{P.~S.} \bibnamefont{Julienne}},
  \bibinfo{author}{\bibfnamefont{D.~S.} \bibnamefont{Jin}}, \bibnamefont{and}
  \bibinfo{author}{\bibfnamefont{J.}~\bibnamefont{Ye}},
  \bibinfo{journal}{Science} \textbf{\bibinfo{volume}{322}},
  \bibinfo{pages}{231} (\bibinfo{year}{2008}).

\bibitem[{\citenamefont{Carr et~al.}(2009)\citenamefont{Carr, DeMille, Krems,
  and Ye}}]{NewJPhys.11.055049}
\bibinfo{author}{\bibfnamefont{L.~D.} \bibnamefont{Carr}},
  \bibinfo{author}{\bibfnamefont{D.}~\bibnamefont{DeMille}},
  \bibinfo{author}{\bibfnamefont{R.~V.} \bibnamefont{Krems}}, \bibnamefont{and}
  \bibinfo{author}{\bibfnamefont{J.}~\bibnamefont{Ye}}, \bibinfo{journal}{New
  J. Phys.} \textbf{\bibinfo{volume}{11}}, \bibinfo{pages}{055049}
  (\bibinfo{year}{2009}).

\bibitem[{\citenamefont{Bahns et~al.}(1996)\citenamefont{Bahns, Stwalley, and
  Gould}}]{JChemPhys.104.9689}
\bibinfo{author}{\bibfnamefont{J.~T.} \bibnamefont{Bahns}},
  \bibinfo{author}{\bibfnamefont{W.~C.} \bibnamefont{Stwalley}},
  \bibnamefont{and} \bibinfo{author}{\bibfnamefont{P.~L.} \bibnamefont{Gould}},
  \bibinfo{journal}{J. Chem. Phys.} \textbf{\bibinfo{volume}{104}},
  \bibinfo{pages}{9689} (\bibinfo{year}{1996}).

\bibitem[{\citenamefont{Di~Rosa}(2004)}]{EurPhysJD.31.395}
\bibinfo{author}{\bibfnamefont{M.~D.} \bibnamefont{Di~Rosa}},
  \bibinfo{journal}{Eur. Phys. J. D} \textbf{\bibinfo{volume}{31}},
  \bibinfo{pages}{395} (\bibinfo{year}{2004}).

\bibitem[{\citenamefont{Stuhl et~al.}(2008)\citenamefont{Stuhl, Sawyer, Wang,
  and Ye}}]{PhysRevLett.101.243002}
\bibinfo{author}{\bibfnamefont{B.~K.} \bibnamefont{Stuhl}},
  \bibinfo{author}{\bibfnamefont{B.~C.} \bibnamefont{Sawyer}},
  \bibinfo{author}{\bibfnamefont{D.}~\bibnamefont{Wang}}, \bibnamefont{and}
  \bibinfo{author}{\bibfnamefont{J.}~\bibnamefont{Ye}}, \bibinfo{journal}{Phys.
  Rev. Lett.} \textbf{\bibinfo{volume}{101}}, \bibinfo{pages}{243002}
  (\bibinfo{year}{2008}).

\bibitem[{\citenamefont{Shuman et~al.}(2009)\citenamefont{Shuman, Barry, Glenn,
  and DeMille}}]{PhysRevLett.103.223001}
\bibinfo{author}{\bibfnamefont{E.~S.} \bibnamefont{Shuman}},
  \bibinfo{author}{\bibfnamefont{J.~F.} \bibnamefont{Barry}},
  \bibinfo{author}{\bibfnamefont{D.~R.} \bibnamefont{Glenn}}, \bibnamefont{and}
  \bibinfo{author}{\bibfnamefont{D.}~\bibnamefont{DeMille}},
  \bibinfo{journal}{Phys. Rev. Lett.} \textbf{\bibinfo{volume}{103}},
  \bibinfo{pages}{223001} (\bibinfo{year}{2009}).

\bibitem[{\citenamefont{Lev et~al.}(2008)\citenamefont{Lev, Vukics, Hudson,
  Sawyer, Domokos, Ritsch, and Ye}}]{PhysRevA.77.023402}
\bibinfo{author}{\bibfnamefont{B.~L.} \bibnamefont{Lev}},
  \bibinfo{author}{\bibfnamefont{A.}~\bibnamefont{Vukics}},
  \bibinfo{author}{\bibfnamefont{E.~R.} \bibnamefont{Hudson}},
  \bibinfo{author}{\bibfnamefont{B.~C.} \bibnamefont{Sawyer}},
  \bibinfo{author}{\bibfnamefont{P.}~\bibnamefont{Domokos}},
  \bibinfo{author}{\bibfnamefont{H.}~\bibnamefont{Ritsch}}, \bibnamefont{and}
  \bibinfo{author}{\bibfnamefont{J.}~\bibnamefont{Ye}}, \bibinfo{journal}{Phys.
  Rev. A} \textbf{\bibinfo{volume}{77}}, \bibinfo{pages}{023402}
  (\bibinfo{year}{2008}).

\bibitem[{\citenamefont{Shuman et~al.}(2010)\citenamefont{Shuman, Barry, and
  DeMille}}]{Nature.467.820}
\bibinfo{author}{\bibfnamefont{E.~S.} \bibnamefont{Shuman}},
  \bibinfo{author}{\bibfnamefont{J.~F.} \bibnamefont{Barry}}, \bibnamefont{and}
  \bibinfo{author}{\bibfnamefont{D.}~\bibnamefont{DeMille}},
  \bibinfo{journal}{Nature} \textbf{\bibinfo{volume}{467}},
  \bibinfo{pages}{820} (\bibinfo{year}{2010}).

\bibitem[{\citenamefont{Nguyen et~al.}()\citenamefont{Nguyen, Viteri,
  Hohenstein, Sherrill, Brown, and Odom}}]{our.paper}
\bibinfo{author}{\bibfnamefont{J.}~\bibnamefont{Nguyen}},
  \bibinfo{author}{\bibfnamefont{C.~R.} \bibnamefont{Viteri}},
  \bibinfo{author}{\bibfnamefont{E.}~\bibnamefont{Hohenstein}},
  \bibinfo{author}{\bibfnamefont{C.~D.} \bibnamefont{Sherrill}},
  \bibinfo{author}{\bibfnamefont{K.~R.} \bibnamefont{Brown}}, \bibnamefont{and}
  \bibinfo{author}{\bibfnamefont{B.}~\bibnamefont{Odom}},
  \bibinfo{note}{arXiv:1102.3368}.

\bibitem[{\citenamefont{Herzberg}(1950)}]{Herzberg}
\bibinfo{author}{\bibfnamefont{G.}~\bibnamefont{Herzberg}},
  \emph{\bibinfo{title}{Molecular Spectra and Molecular Structure: I. Spectra
  of Diatomic Molecules}}, Molecular Spectra and Molecular Structure
  (\bibinfo{publisher}{Van Nostrand Reinhold Company}, \bibinfo{address}{New
  York}, \bibinfo{year}{1950}), \bibinfo{edition}{2nd} ed.

\bibitem[{\citenamefont{Cai and François}(1999)}]{JMolSpec.197.12}
\bibinfo{author}{\bibfnamefont{Z.~L.} \bibnamefont{Cai}} \bibnamefont{and}
  \bibinfo{author}{\bibfnamefont{J.~P.} \bibnamefont{François}},
  \bibinfo{journal}{J. Mol. Spectrosc.} \textbf{\bibinfo{volume}{197}},
  \bibinfo{pages}{12 } (\bibinfo{year}{1999}).

\bibitem[{\citenamefont{Viteau et~al.}(2008)\citenamefont{Viteau, Chotia,
  Allegrini, Bouloufa, Dulieu, Comparat, and Pillet}}]{Science.321.232}
\bibinfo{author}{\bibfnamefont{M.}~\bibnamefont{Viteau}},
  \bibinfo{author}{\bibfnamefont{A.}~\bibnamefont{Chotia}},
  \bibinfo{author}{\bibfnamefont{M.}~\bibnamefont{Allegrini}},
  \bibinfo{author}{\bibfnamefont{N.}~\bibnamefont{Bouloufa}},
  \bibinfo{author}{\bibfnamefont{O.}~\bibnamefont{Dulieu}},
  \bibinfo{author}{\bibfnamefont{D.}~\bibnamefont{Comparat}}, \bibnamefont{and}
  \bibinfo{author}{\bibfnamefont{P.}~\bibnamefont{Pillet}},
  \bibinfo{journal}{Science} \textbf{\bibinfo{volume}{321}},
  \bibinfo{pages}{232} (\bibinfo{year}{2008}).

\bibitem[{\citenamefont{Mogi et~al.}(2002)\citenamefont{Mogi, Fukuyama,
  Kobayashi, Tanihata, Uehara, and Matsuo}}]{AppSurfSci.197.202}
\bibinfo{author}{\bibfnamefont{T.}~\bibnamefont{Mogi}},
  \bibinfo{author}{\bibfnamefont{Y.}~\bibnamefont{Fukuyama}},
  \bibinfo{author}{\bibfnamefont{T.}~\bibnamefont{Kobayashi}},
  \bibinfo{author}{\bibfnamefont{I.}~\bibnamefont{Tanihata}},
  \bibinfo{author}{\bibfnamefont{K.}~\bibnamefont{Uehara}}, \bibnamefont{and}
  \bibinfo{author}{\bibfnamefont{Y.}~\bibnamefont{Matsuo}},
  \bibinfo{journal}{Appl. Surf. Sci.} \textbf{\bibinfo{volume}{197-198}},
  \bibinfo{pages}{202 } (\bibinfo{year}{2002}).

\bibitem[{\citenamefont{Staanum et~al.}(2010)\citenamefont{Staanum, Hojbjerre,
  Skyt, Hansen, and Drewsen}}]{NatPhys.6.271}
\bibinfo{author}{\bibfnamefont{P.~F.} \bibnamefont{Staanum}},
  \bibinfo{author}{\bibfnamefont{K.}~\bibnamefont{Hojbjerre}},
  \bibinfo{author}{\bibfnamefont{P.~S.} \bibnamefont{Skyt}},
  \bibinfo{author}{\bibfnamefont{A.~K.} \bibnamefont{Hansen}},
  \bibnamefont{and} \bibinfo{author}{\bibfnamefont{M.}~\bibnamefont{Drewsen}},
  \bibinfo{journal}{Nat. Phys.} \textbf{\bibinfo{volume}{6}},
  \bibinfo{pages}{271} (\bibinfo{year}{2010}).

\bibitem[{\citenamefont{Schneider et~al.}(2010)\citenamefont{Schneider, Roth,
  Duncker, Ernsting, and Schiller}}]{NatPhys.6.275}
\bibinfo{author}{\bibfnamefont{T.}~\bibnamefont{Schneider}},
  \bibinfo{author}{\bibfnamefont{B.}~\bibnamefont{Roth}},
  \bibinfo{author}{\bibfnamefont{H.}~\bibnamefont{Duncker}},
  \bibinfo{author}{\bibfnamefont{I.}~\bibnamefont{Ernsting}}, \bibnamefont{and}
  \bibinfo{author}{\bibfnamefont{S.}~\bibnamefont{Schiller}},
  \bibinfo{journal}{Nat. Phys.} \textbf{\bibinfo{volume}{6}},
  \bibinfo{pages}{275} (\bibinfo{year}{2010}).

\bibitem[{\citenamefont{Le~Roy}(|2007|)}]{level8.0}
\bibinfo{author}{\bibfnamefont{R.~J.} \bibnamefont{Le~Roy}},
  \bibinfo{type}{Chemical Physics Research Report} \bibinfo{number}{CP-663},
  \bibinfo{institution}{University of Waterloo} (\bibinfo{year}{|2007|}),
  \bibinfo{note}{the source code and manual for this program may be obtained
  from http://leroy.uwaterloo.ca/programs/.}

\bibitem[{\citenamefont{Cameron et~al.}(1995)\citenamefont{Cameron, Scholl,
  Zhang, Holt, and Rosner}}]{JMolSpec.169.364}
\bibinfo{author}{\bibfnamefont{R.}~\bibnamefont{Cameron}},
  \bibinfo{author}{\bibfnamefont{T.~J.} \bibnamefont{Scholl}},
  \bibinfo{author}{\bibfnamefont{L.}~\bibnamefont{Zhang}},
  \bibinfo{author}{\bibfnamefont{R.~A.} \bibnamefont{Holt}}, \bibnamefont{and}
  \bibinfo{author}{\bibfnamefont{S.~D.} \bibnamefont{Rosner}},
  \bibinfo{journal}{J. Mol. Spectrosc.} \textbf{\bibinfo{volume}{169}},
  \bibinfo{pages}{364 } (\bibinfo{year}{1995}).

\bibitem[{\citenamefont{Werner et~al.}(1982)\citenamefont{Werner, Rosmus, and
  Grimm}}]{ChemPhys.73.169}
\bibinfo{author}{\bibfnamefont{H.-J.} \bibnamefont{Werner}},
  \bibinfo{author}{\bibfnamefont{P.}~\bibnamefont{Rosmus}}, \bibnamefont{and}
  \bibinfo{author}{\bibfnamefont{M.}~\bibnamefont{Grimm}},
  \bibinfo{journal}{Chem. Phys.} \textbf{\bibinfo{volume}{73}},
  \bibinfo{pages}{169 } (\bibinfo{year}{1982}).

\bibitem[{\citenamefont{Marian et~al.}(1989)\citenamefont{Marian, Larsson,
  Olsson, and Sigray}}]{ChemPhys.130.361}
\bibinfo{author}{\bibfnamefont{C.}~\bibnamefont{Marian}},
  \bibinfo{author}{\bibfnamefont{M.}~\bibnamefont{Larsson}},
  \bibinfo{author}{\bibfnamefont{B.}~\bibnamefont{Olsson}}, \bibnamefont{and}
  \bibinfo{author}{\bibfnamefont{P.}~\bibnamefont{Sigray}},
  \bibinfo{journal}{Chem. Phys.} \textbf{\bibinfo{volume}{130}},
  \bibinfo{pages}{361 } (\bibinfo{year}{1989}).

\bibitem[{\citenamefont{Clouthier and Grein}(2005)}]{ChemPhys.315.35}
\bibinfo{author}{\bibfnamefont{C.~M.} \bibnamefont{Clouthier}}
  \bibnamefont{and} \bibinfo{author}{\bibfnamefont{F.}~\bibnamefont{Grein}},
  \bibinfo{journal}{Chem. Phys.} \textbf{\bibinfo{volume}{315}},
  \bibinfo{pages}{35 } (\bibinfo{year}{2005}).

\bibitem[{\citenamefont{Giri and Das}(2003)}]{ChemPhysLett.368.465}
\bibinfo{author}{\bibfnamefont{D.}~\bibnamefont{Giri}} \bibnamefont{and}
  \bibinfo{author}{\bibfnamefont{K.~K.} \bibnamefont{Das}},
  \bibinfo{journal}{Chem. Phys. Lett.} \textbf{\bibinfo{volume}{368}},
  \bibinfo{pages}{465 } (\bibinfo{year}{2003}).

\bibitem[{\citenamefont{Balasubramanian}(1984)}]{JPhysChem.88.5759}
\bibinfo{author}{\bibfnamefont{K.}~\bibnamefont{Balasubramanian}},
  \bibinfo{journal}{J. Phys. Chem.} \textbf{\bibinfo{volume}{88}},
  \bibinfo{pages}{5759} (\bibinfo{year}{1984}).

\bibitem[{\citenamefont{Wang et~al.}(2009)\citenamefont{Wang, Yang, Su, Bai,
  and Wang}}]{ChinOptLett.8.663}
\bibinfo{author}{\bibfnamefont{X.}~\bibnamefont{Wang}},
  \bibinfo{author}{\bibfnamefont{C.}~\bibnamefont{Yang}},
  \bibinfo{author}{\bibfnamefont{T.}~\bibnamefont{Su}},
  \bibinfo{author}{\bibfnamefont{F.}~\bibnamefont{Bai}}, \bibnamefont{and}
  \bibinfo{author}{\bibfnamefont{M.}~\bibnamefont{Wang}},
  \bibinfo{journal}{Chin. Opt. Lett.} \textbf{\bibinfo{volume}{7}},
  \bibinfo{pages}{663} (\bibinfo{year}{2009}).

\bibitem[{\citenamefont{Zenouda et~al.}(1998)\citenamefont{Zenouda, Blottiau,
  Chambaud, and Rosmus}}]{JMolStruct.458.61}
\bibinfo{author}{\bibfnamefont{C.}~\bibnamefont{Zenouda}},
  \bibinfo{author}{\bibfnamefont{P.}~\bibnamefont{Blottiau}},
  \bibinfo{author}{\bibfnamefont{G.}~\bibnamefont{Chambaud}}, \bibnamefont{and}
  \bibinfo{author}{\bibfnamefont{P.}~\bibnamefont{Rosmus}},
  \bibinfo{journal}{J. Mol. Struct.} \textbf{\bibinfo{volume}{458}},
  \bibinfo{pages}{61 } (\bibinfo{year}{1998}).

\bibitem[{\citenamefont{Petsalakis et~al.}(2004)\citenamefont{Petsalakis,
  Theodorakopoulos, Gora, and Roszak}}]{JMolStruct.672.105}
\bibinfo{author}{\bibfnamefont{I.~D.} \bibnamefont{Petsalakis}},
  \bibinfo{author}{\bibfnamefont{G.}~\bibnamefont{Theodorakopoulos}},
  \bibinfo{author}{\bibfnamefont{R.~W.} \bibnamefont{Gora}}, \bibnamefont{and}
  \bibinfo{author}{\bibfnamefont{S.}~\bibnamefont{Roszak}},
  \bibinfo{journal}{J. Mol. Struct.} \textbf{\bibinfo{volume}{672}},
  \bibinfo{pages}{105 } (\bibinfo{year}{2004}).

\bibitem[{\citenamefont{Balfour et~al.}(1996)\citenamefont{Balfour, Saksena,
  Shetty, Brown, Barrow, Malcolm, James, and Simard}}]{MolPhys.89.13}
\bibinfo{author}{\bibfnamefont{W.~J.} \bibnamefont{Balfour}},
  \bibinfo{author}{\bibfnamefont{M.~D.} \bibnamefont{Saksena}},
  \bibinfo{author}{\bibfnamefont{B.~J.} \bibnamefont{Shetty}},
  \bibinfo{author}{\bibfnamefont{J.~M.} \bibnamefont{Brown}},
  \bibinfo{author}{\bibfnamefont{R.~F.} \bibnamefont{Barrow}},
  \bibinfo{author}{\bibfnamefont{I.~B.} \bibnamefont{Malcolm}},
  \bibinfo{author}{\bibfnamefont{A.~M.} \bibnamefont{James}}, \bibnamefont{and}
  \bibinfo{author}{\bibfnamefont{B.}~\bibnamefont{Simard}},
  \bibinfo{journal}{Mol. Phys.} \textbf{\bibinfo{volume}{89}},
  \bibinfo{pages}{13 } (\bibinfo{year}{1996}).

\bibitem[{\citenamefont{Langhoff et~al.}(1987)\citenamefont{Langhoff, Charles
  W.~Bauschlicher, and Partridge}}]{JChemPhys.87.4716}
\bibinfo{author}{\bibfnamefont{S.~R.} \bibnamefont{Langhoff}},
  \bibinfo{author}{\bibfnamefont{J.}~\bibnamefont{Charles W.~Bauschlicher}},
  \bibnamefont{and}
  \bibinfo{author}{\bibfnamefont{H.}~\bibnamefont{Partridge}},
  \bibinfo{journal}{J. Chem. Phys.} \textbf{\bibinfo{volume}{87}},
  \bibinfo{pages}{4716} (\bibinfo{year}{1987}).

\bibitem[{\citenamefont{Langhoff and Charles
  W.~Bauschlicher}(1988)}]{JChemPhys.88.329}
\bibinfo{author}{\bibfnamefont{S.~R.} \bibnamefont{Langhoff}} \bibnamefont{and}
  \bibinfo{author}{\bibfnamefont{J.}~\bibnamefont{Charles W.~Bauschlicher}},
  \bibinfo{journal}{J. Chem. Phys.} \textbf{\bibinfo{volume}{88}},
  \bibinfo{pages}{329} (\bibinfo{year}{1988}).

\bibitem[{\citenamefont{Dubois and Lefèbvre}(2004)}]{MolPhys.102.23}
\bibinfo{author}{\bibfnamefont{I.}~\bibnamefont{Dubois}} \bibnamefont{and}
  \bibinfo{author}{\bibfnamefont{P.-H.} \bibnamefont{Lefèbvre}},
  \bibinfo{journal}{Mol. Phys.} \textbf{\bibinfo{volume}{102}},
  \bibinfo{pages}{23 } (\bibinfo{year}{2004}).

\bibitem[{\citenamefont{Sannigrahi et~al.}(1995)\citenamefont{Sannigrahi,
  Buenker, Hirsch, and Gu}}]{ChemPhsLett.237.204}
\bibinfo{author}{\bibfnamefont{A.~B.} \bibnamefont{Sannigrahi}},
  \bibinfo{author}{\bibfnamefont{R.~J.} \bibnamefont{Buenker}},
  \bibinfo{author}{\bibfnamefont{G.}~\bibnamefont{Hirsch}}, \bibnamefont{and}
  \bibinfo{author}{\bibfnamefont{J.-P.} \bibnamefont{Gu}},
  \bibinfo{journal}{Chem. Phys. Lett.} \textbf{\bibinfo{volume}{237}},
  \bibinfo{pages}{204 } (\bibinfo{year}{1995}).

\bibitem[{\citenamefont{Nishimura et~al.}(1983)\citenamefont{Nishimura,
  Mizuguchi, Tsuji, Obara, and Morokuma}}]{JChemPhys.78.7260}
\bibinfo{author}{\bibfnamefont{Y.}~\bibnamefont{Nishimura}},
  \bibinfo{author}{\bibfnamefont{T.}~\bibnamefont{Mizuguchi}},
  \bibinfo{author}{\bibfnamefont{M.}~\bibnamefont{Tsuji}},
  \bibinfo{author}{\bibfnamefont{S.}~\bibnamefont{Obara}}, \bibnamefont{and}
  \bibinfo{author}{\bibfnamefont{K.}~\bibnamefont{Morokuma}},
  \bibinfo{journal}{J. Chem. Phys.} \textbf{\bibinfo{volume}{78}},
  \bibinfo{pages}{7260} (\bibinfo{year}{1983}).

\bibitem[{\citenamefont{Tong et~al.}(2010)\citenamefont{Tong, Winney, and
  Willitsch}}]{PhysRevLett.105.143001}
\bibinfo{author}{\bibfnamefont{X.}~\bibnamefont{Tong}},
  \bibinfo{author}{\bibfnamefont{A.~H.} \bibnamefont{Winney}},
  \bibnamefont{and}
  \bibinfo{author}{\bibfnamefont{S.}~\bibnamefont{Willitsch}},
  \bibinfo{journal}{Phys. Rev. Lett.} \textbf{\bibinfo{volume}{105}},
  \bibinfo{pages}{143001} (\bibinfo{year}{2010}).

\end{thebibliography}

\end{document}